\documentclass[12pt,showpacs,preprint,nofootinbib,showkeys]{revtex4}
\usepackage[intlimits]{amsmath}
\usepackage{amssymb}
\usepackage{exscale}
\usepackage{graphicx}
\usepackage{setspace}
\usepackage{mathrsfs}

\newcommand{\ud}{\text{d}}
\newcommand{\im}{\mathrm{i}}
\newcommand{\T}{\mathrm{T}}
\newcommand{\tr}{\mathrm{tr}}
\newcommand{\CL}{\mathscr{L}}

\allowdisplaybreaks

\begin{document}

\vspace{1 cm}

\title{Existence of spinning solitons in gauge field theory}

\author{Mikhail S. Volkov\footnotemark[2]}
\author{Erik W\"ohnert\footnotemark[3]}

\vspace{1 cm}

\affiliation{\footnotemark[2]Laboratoire de Math\'ematiques et Physique
Th\'eorique, Universit\'e de Tours, Parc de Grandmont, 37200 Tours,
FRANCE \footnotetext[2]{\tt volkov@phys.univ-tours.fr} \\
\footnotemark[3]Theoretisch-Physikalisches Institut, Friedrich
Schiller Universit\"at Jena, Fr\"obelstieg 1, 07743 Jena,
GERMANY \footnotetext[3]{\tt pew@tpi.uni-jena.de}}


\begin{abstract}

\noindent 
We study the existence of classical soliton solutions with intrinsic
angular momentum in Yang-Mills-Higgs theory with a compact gauge
group $\mathcal{G}$ in ($3+1$)-dimensional Minkowski space. 
We show that for \textit{symmetric} gauge fields the Noether charges 
corresponding to \textit{rigid} spatial symmetries, as the angular momentum, 
can be expressed in terms of \textit{surface}
integrals. Using this result, we demonstrate 
in the case of $\mathcal{G}=SU(2)$
the nonexistence of stationary
and axially symmetric spinning excitations
for all known topological solitons in the one-soliton sector, 
that is, for 't~Hooft--Polyakov
monopoles, Julia-Zee dyons, sphalerons, and also vortices.

\end{abstract}

\maketitle


\section{Introduction}

The existence of globally regular soliton solutions with a nonvanishing
angular momentum in classical field theory is an interesting open issue,
which has recently been addressed in a number of publications
\cite{Schunk96,Yoshida:1997qf,Heusler:1998ec,VanderBij:2001nm,%
Volkov:2002aj,VanderBij:2002sq,Radu:2002rv}.  Up to now, such spinning
solutions in Minkowski space have been found only in the theory of a
self-interacting complex scalar field (Q-balls)
\cite{Volkov:2002aj}\footnote{In curved space similar rotating solutions
are known for a self-gravitating scalar field (boson stars)
\cite{Schunk96,Yoshida:1997qf}.}. For these solutions the
energy-momentum tensor is stationary and axially symmetric, while the
angular momentum $J \sim \omega N$ is generated by the rotating phase of
the scalar field $\Phi = \phi(r,\vartheta) e^{-\im \omega t + \im N
\varphi}$.

It is natural to wonder whether rotating solitons can also exist in
gauge field theories with spontaneously broken symmetries.  For
stationary, axially symmetric systems the rotating phase of the Higgs
field can be gauged away\footnote{It is not excluded that the action
could be invariant under time translations and axial rotations while the
fields are not stationary and axially symmetric. In such a case it would
not be possible to gauge away the rotating phases. Such a possibility,
however, is beyond the scope of our present consideration.}. A nonzero
angular momentum could then be supported only by the Poynting vector of
the gauge field, and in fact such solutions can
indeed be obtained, as for example rotating 
\cite{Heusler:1998ec,VanderBij:2001nm} monopole-antimonopole pairs
\cite{Kleihaus:1999sx,Taubes:1982ie,Taubes:1982if}. However, the
rotation is then rather associated with the orbital motion in a
many-body system. The real challenge is to construct rotating solutions
in the \textit{one}-soliton sector, where the rotation would indeed be
associated with spinning excitations of an individual object. For some
strange reason, up to now such spinning solitons have been found only in
anti--de Sitter space \cite{VanderBij:2002sq,Radu:2002rv}, while their
possible existence in Minkowski space remains rather obscure. In fact,
the results obtained so far in this area have all been negative. For
example, it has been shown that 't~Hooft--Polyakov monopoles and 
Julia-Zee dyons in Minkowski space cannot rotate \textit{slowly}
\cite{Heusler:1998ec}. The same conclusion holds for
gravitating monopoles and sphalerons \cite{Brodbeck:1997sa}. In
addition, it was noticed in \cite{VanderBij:2001nm} that for
axially symmetric deformations of
Julia-Zee dyons the angular momentum can be
represented as a flux integral, a fact that was used 
in \cite{VanderBij:2001nm}  to argue that 
dyons cannot rotate \textit{rapidly} either.  

In this paper we study the existence of spinning solitons in the context
of Yang-Mills-Higgs (YMH) theory for an arbitrary compact gauge
group $\mathcal{G}$ in $(3+1)$-dimensional Minkowski space. First of all,
we analyze the observation of \cite{VanderBij:2001nm} that the angular
momentum of Julia-Zee dyons can be expressed as a flux integral. It
is natural to wonder why such a representation of the angular momentum
exists at all and whether it can be generalized to other models.
Usually, conserved quantities associated with Poincar\'e symmetries in
Minkowski space, such as, for example, the energy, are given by volume
integrals and not surface integrals. We therefore study the
relationship between conservation laws, spacetime, and gauge symmetries, and
what we find is the following. For \textit{symmetric} gauge
fields, the action of a \textit{rigid} spacetime symmetry 
generated by a Killing vector $X$ is equivalent
to that of a \textit{local} gauge symmetry generated by a Lie algebra 
valued function $W_X$. It is then a consequence of
the Bianchi identities imposed by the local gauge symmetry that the
Noether current for the global Poincar\'e symmetry is essentially a
total divergence. In the case of \textit{spatial}
symmetries, the Noether charge can then be expressed by
a surface integral 
\begin{align}
\label{Main)}
\Theta_{X} & = \oint \langle (A_{X}-W_X) F^{0k} \rangle \ud S_k \;,
\end{align}
where $A_X$ is the $X$ projection of the gauge field. In the case of 
spatial rotations  $X=\frac{\partial}{\partial\varphi}$,
 this gives the conserved and gauge invariant angular momentum. 

Making use of this representation of
the angular momentum, we then systematically study the fields in
the asymptotic region near spatial infinity, looking for field modes 
that could give a contribution to the surface integral.
In this way we show that for 't~Hooft--Polyakov monopoles and Julia-Zee
dyons there are no stationary, axially symmetric deformations giving a
nonzero contribution to the angular momentum.  We then carry out a
similar analysis for sphalerons and also for vortices --- with the same
conclusion. As a result, we in fact show the absence of stationary and
axially symmetric spinning excitations in the one-soliton sector
for all known topological
solitons with gauge group $\mathcal{G} = SU(2)$. 
The still remaining
possibilities of constructing rotating solutions can then be
related only either to studying solutions with higher gauge groups or to
considering fields that are not manifestly stationary or axially
symmetric.

\section{Yang-Mills-Higgs theory}

The theory under consideration is a Yang-Mills-Higgs theory with compact
gauge group $\mathcal{G}$ defined by the action
\begin{align}
S_\textrm{YMH} & = \int \CL \, \ud^4 x \;,
\end{align}
where
\begin{align}
\CL & = -\frac{1} {4} \langle F_{\mu\nu} F^{\mu\nu} \rangle
+ \frac{1} {2} \, (\mathcal{D}_{\mu} \Phi)^{\dagger} \mathcal{D}^{\mu}
\Phi - \frac{\lambda} {4} \,(\Phi^{\dagger} \Phi - 1)^2 \, .
\end{align}
Here, the gauge field strength $F_{\mu\nu} \equiv {\T}_a F^a_{~\mu\nu}
= \partial_{\mu} A_{\nu} - \partial_{\nu} A_{\mu} + [A_{\mu} , A_{\nu}]$
with the gauge field $A_{\mu} \equiv {\T}_{a} A^{a}_{\mu} $. The
anti-Hermitian gauge group generators ${\T}_{a}$ ($a = 1,2, \ldots,
\dim \mathcal{G}$) satisfy the relations
\begin{alignat}{2}
[\T_{a}, \T_{b}] & = f_{abc} \T_c \;, & \quad \tr (\T_{a} \T_{b}) & = K
\delta_{ab} \;.
\end{alignat}
The invariant scalar product in the Lie algebra is defined as 
$\langle A B \rangle = \frac{1} {K} \tr (A B)$. The Higgs field $\Phi$
is a vector in the representation space of $\mathcal{G}$ where the
generators $\T_a$ act; this space can be complex or
real. $\mathcal{D}_{\mu} \Phi = (\partial_{\mu} + A_{\mu}) \Phi$ is the
covariant derivative of the Higgs field. The units are chosen such that
the gauge coupling constant and the vacuum value of the Higgs field are
equal to 1. Spacetime indices are lifted with the Minkowski metric.

Below, we will consider two important particular cases corresponding to
$\mathcal{G} = SU(2)$. The Higgs field $\Phi$ then will be chosen to be
either in the real triplet representation, in which case
\begin{align}
\label{triplet}
(\T_{a})_{ik} & = - \varepsilon_{aik} \;,
\end{align}
or in the complex doublet representation with 
\begin{align}
\label{dublet}
\T_{a} & = \frac{\tau_a} {2 i} \;,
\end{align}
where $\tau_a$ are the Pauli matrices. 

The action is invariant under gauge transformations
\begin{alignat}{2}
\label{3}
A_{\mu} & \to U (A_{\mu} + \partial_{\mu}) U^{-1} \;, & \quad
\Phi & \to U \Phi \;,
\end{alignat}
where $U$ is a $\mathcal{G}$ valued function. Varying the action with
respect to the gauge and Higgs fields gives the equations of motion 
\begin{align}
%
%
\label{eq1}
D_{\sigma} F^{\sigma\mu} & = \frac{1} {2} \big( \Phi^{\dagger} \T_{a}
\mathcal{D}^{\mu} \Phi -(\mathcal{D}^{\mu} \Phi)^{\dagger} \T_{a} \Phi
\big) \T_{a} \;, \\
%
%
\label{eq2}
\mathcal{D}_{\mu} \mathcal{D}^{\mu} \Phi & = - \lambda (\Phi^{\dagger}
\Phi - 1) \Phi \;,
\end{align}
where $D_{\mu} = \partial_{\mu} + [A_{\mu},~~]$ is the covariant
derivative in the adjoint representation. 

In what follows, we will consider stationary, axially symmetric fields
subject to the symmetry conditions \cite{Forgacs:1980zs} 
\begin{alignat}{3}
\label{symmetry}
\mathcal{L}_{\xi_m} A_{\mu} & = D_{\mu} W_{m} \;, & \quad
\mathcal{L}_{\xi_m} \Phi & = - W_{m} \Phi \;, & \quad m & = t,\varphi \;.
\end{alignat}
Here, $\mathcal{L}_{\xi_m}$ are the Lie derivatives along the two Killing
vectors  $\xi_{t} = {\partial_t}$ 
and  $\xi_{\varphi} = {\partial_{\varphi}}$, 
while $W_m$ are compensating Lie algebra valued functions. The general
solution of these equations is well known: since the two Killing vectors
commute, there exists a gauge where $W_m = 0$. Therefore, the symmetry
conditions in this gauge require simply the independence from $t$ and
$\varphi$. As a result, the most general solution is
\begin{alignat}{2}
\label{s}
A_{\mu} & = \T_{a} \, A^{a}_{\mu} (\rho,z) \, \ud x^\mu \;, & \quad
\Phi & = \Phi(\rho,z) \;.
\end{alignat}
The regularity on the symmetry axis requires that 
\begin{align}
\label{axe}
A_{\varphi}(0,z) & = f(z) \, \T \;,
\end{align}
where $\T$ is an element of the Cartan subalgebra of the Lie algebra of
$\mathcal{G}$ and $f(z)$ is a bounded function. Passing to a new gauge
with the gauge transformation $U = e^{-\varphi f(z) \T}$ will then send
$A_\varphi(0,z)$ to zero.

\section{Noether charges as flux integrals}

Conserved quantities in field theory are determined by Noether charges
corresponding to global symmetries of the action. These charges can be
expressed as volume integrals of the local charge densities. In some
cases, such as, for example, for the electric charge, these volume integrals
can be further transformed to surface integrals. The reason for this is
as follows (see \cite{Julia:1998ys} for a discussion). 
Electric charge is conserved owing to the invariance under
global phase rotations. In gauge field theory, this symmetry is a
special case of the local gauge invariance.  The local gauge invariance
leads to the existence of identity relations between the field equations
(Bianchi identities) and implies the identical conservation of Noether's
currents, since they can be represented as divergences of antisymmetric
quantities (sometimes called superpotentials)
\begin{align}
\label{local}
\Theta^{\mu} & = \partial_{\sigma} (\omega(x) \mathcal{F}^{\sigma \mu})
\;. 
\end{align}
Here, $\omega(x)$ is the parameter of local gauge transformations, the
case of global phase rotations corresponding to constant
$\omega$'s. Since $\Theta^0$ is a total divergence, the Noether charge
can be expressed as a surface integral.

The procedure described above is very well illustrated in the context of
general relativity, where the conserved energy, momentum, and angular
momentum are given by flux integrals. 
This can be traced back to the
fact that the Poincar\'e symmetries are a special case of general
spacetime diffeomorphisms. 
For theories in Minkowski space, on the other
hand, there is no diffeomorphism invariance, and so Poincar\'e symmetries
are not related to any local symmetries. As a result, the energy, for
example, cannot be expressed as a flux integral. However, for
\textit{symmetric} gauge fields some of the spacetime symmetries can be
equivalent to local gauge symmetries in the sense that the result of
Poincar\'e transformations can be compensated by gauge transformations.
As a result, the corresponding Noether charges will have essentially the
same structure as in Eq.(\ref{local}), and the Noether charges can be
expressed as flux integrals. We will now show how this works in the
context of YMH theory.

It is well known \cite{Jackiw:1978ar,Jackiw:1980ub} that in the presence of
gauge invariance spacetime symmetries must be combined with the internal
gauge symmetries in order to give conserved and \textit{gauge invariant} 
charges
via Noether's procedure.  If $X^\mu$ is a Killing vector of the
system\footnote{Thus, one has $\partial_\mu X_\nu+\partial_\nu
X_\mu=0$.} then the corresponding conserved and gauge invariant Noether
current is
\begin{align}
\label{Noether}
\Theta^{\mu} & = \sum_{B} \frac{\partial \CL} {\partial(\partial_{\mu}
u^{B})} \, \delta u^{B} - X^{\mu} \CL \;.
\end{align}
Here, $u^B$ collectively denotes the fields $(A_{\mu}, \Phi,
\Phi^{\dagger})$, and the variations $\delta u^B$ include the part
generated by $X^\mu$ plus another part due to an infinitesimal
gauge transformation generated by a Lie algebra valued function $W$:
\begin{align}
\label{delta}
\delta u^{B} & = \mathcal{L}_{X} u^{B} - \delta_{W} u^{B} \;.
\end{align}
Here, the Lie derivatives are 
\begin{align}
\label{Lie}
\mathcal{L}_{X} A_{\mu} & = X^{\alpha} \partial_{\alpha} A_{\mu} +
A^{\alpha} \partial_{\alpha} X_{\mu} \;, &
\mathcal{L}_{X} \Phi & = X^{\alpha} \partial_{\alpha} \Phi \;, 
\end{align}
while the gauge variations are given by
\begin{align}
\label{gauge}
\delta_{W} A_{\mu} & = D_{\mu} W \;, &
\delta_{W} \Phi & = - W \Phi \;. 
\end{align}
The function $W$ is determined by the requirement that the variations 
$\delta u^{B}$ transform under gauge transformations \textit{covariantly}, thus
ensuring the gauge invariance of the Noether current. Using the identity
\cite{Jackiw:1978ar,Jackiw:1980ub}
\begin{align}
\label{id}
\mathcal{L}_{X} A_{\mu} & = X^{\alpha} F_{\alpha \mu} + D_{\mu}
(X^{\alpha} A_{\alpha}) \;,
\end{align}
one obtains 
\begin{align}                   \label{vvar}
\delta A_{\mu} & = X^{\alpha} F_{\alpha\mu} + D_{\mu} (X^{\alpha}
A_{\alpha} - W) \;, & 
\delta \Phi & = X^{\alpha} \mathcal{D}_{\alpha} \Phi - (X^{\alpha}
A_{\alpha} - W) \;,
\end{align}
which shows that the transformation law for $W$ must be 
\begin{align}     \label{trans}
W & \to U (W + X^{\sigma} \partial_{\sigma}) U^{-1} \;,
\end{align}
since then $(X^{\alpha} A_{\alpha} - W)$ transforms covariantly. 
Having this in mind and 
inserting Eqs.(\ref{delta})--(\ref{gauge}) into Eq.(\ref{Noether}), 
one obtains 
after straightforward calculations 
\begin{align}
\label{Noether4}
\Theta^{\mu} & = X^{\alpha} {T}^{\mu}_{\alpha} + \partial_{\sigma}
\langle (X^{\nu} A_{\nu} - W) F^{\sigma \mu} \rangle \;. 
\end{align}
Here the tensor 
\begin{align}                              \label{Tm}
T^{\mu}_{\nu} & = - \langle F^{\mu\sigma} F_{\nu\sigma} \rangle +
\frac{1} {2} \big((\mathcal{D}^{\mu} \Phi)^{\dagger} \mathcal{D}_{\nu}
\Phi + (\mathcal{D}_{\nu} \Phi)^{\dagger} \mathcal{D}^{\mu} \Phi \big) -
\delta^{\mu}_{\nu} \CL \; 
\end{align}
coincides 
with the metrical energy-momentum tensor obtained by varying the action
with respect to the spacetime metric. This tensor is symmetric and 
divergence-free, $\partial_{\mu} T^{\mu \nu} = 0$. 

The Noether current (\ref{Noether4}) is conserved and gauge invariant. 
However, it is not yet completely defined, since $W$ is not uniquely
determined by the condition (\ref{trans}). This  
reflects the well-known
ambiguity in the definition of Noether currents, as they can 
always be changed by 
adding the divergence of an antisymmetric tensor. 
The way to uniquely define the Noether currents 
(see, for example, \cite{Landau,Jackiw:1978ar,Jackiw:1980ub})
is dictated by the agreement with the general relativity (GR), 
since they should coincide with the conserved currents 
obtained from the metrical energy-momentum tensor. 
The {\it canonical} Noether energy-momentum tensor 
will then be symmetric and will 
coincide with the metrical one.  
All this is achieved if only one chooses \cite{Jackiw:1978ar,Jackiw:1980ub}
\begin{align}
\label{WWW}
W & = X^{\alpha} A_{\alpha}
\end{align}
(notice that this transforms according Eq.(\ref{trans})) in order 
to get rid of the second term on the right in 
Eq.(\ref{Noether4}). The Noether current then becomes 
\begin{align}
\label{Noether5}                 
\Theta^{\mu} & = 
X^{\alpha} {T}^{\mu}_{\alpha}. 
\end{align}
This coincides with the standard GR current 
and leads to the conserved charge expressed by the volume 
integral over the three-space, 
\begin{align}
\label{M}
\Theta_{X} & = \int X^\alpha T^0_\alpha d^3x\;. 
\end{align}
This formula reproduces the known result for the conserved and 
gauge invariant Noether charge associated with a rigid Poincar\'e
symmetry $X^\mu$
\cite{Jackiw:1978ar}.

Let us now repeat the calculation above by 
assuming that the symmetry generated by $X^{\mu}$ is not only
a symmetry of the \textit{action}, but also a symmetry of the
\textit{fields}, in the sense that there exists a Lie algebra valued
function $W_{X}$ such that
\begin{alignat}{2}
\label{symX}
\mathcal{L}_{X} A_{\mu} & = D_{\mu} W_{X} \;, & \quad
\mathcal{L}_{X} \Phi & = - W_{X} \Phi \;.
\end{alignat}
Substituting this into Eq.(\ref{delta}) and using Eq.(\ref{gauge}) gives
\begin{alignat}{2}
\label{var}
\delta A_{\mu} & = -D_{\mu} \Psi_{X} \;, & \quad
\delta \Phi & = \Psi_{X} \Phi \;,
\end{alignat}
where $\Psi_{X} = W - W_{X}=X^\alpha A_\alpha - W_{X} $. 
Therefore, the field variations generated
by $X^{\mu}$ can in this case be viewed as {\it pure gauge} 
variations. Inserting
Eq.(\ref{var}) into Eq.(\ref{Noether}) gives
\begin{align}
\Theta^{\mu} & = \langle F^{\mu\alpha} D_{\alpha} \Psi_{X} \rangle
+ \frac{1} {2} (\mathcal{D}^{\mu} \Phi)^{\dagger} \Psi_{X} \Phi
- \frac{1} {2} \Phi^{\dagger} \Psi_{X} \mathcal{D}^{\mu} \Phi - X^{\mu}
\CL \;, 
\end{align}
and using the equations of motion (\ref{eq1}) this reduces to
\begin{align}
\label{theta}
\Theta^{\mu} & = -\partial_{\alpha} \langle \Psi_{X} F^{\alpha \mu}
\rangle - X^{\mu} \CL \;.
\end{align}
This almost has the structure of an identically conserved current, if it
were not for the last term. This term is the remnant of the fact that the
symmetries under consideration, although closely related to gauge
symmetries, are actually spacetime symmetries. 
Now, if the vector $X^\mu$ is \textit{spacelike},  as is the case
for strictly spatial translations and rotations, then there exist
reference frames where the temporal component $X^0$ vanishes. 
As a result, $\Theta^{0}$ is a
total divergence and its integral 
over the spatial hypersurface
can be transformed into a surface integral
(provided that there is no contribution from the inner boundary). 
The conserved and gauge invariant Noether charge is then given by the
flux integral over a closed two-surface at spatial infinity:
\begin{align}
\label{Main}
\Theta_{X} & = - \oint \langle \Psi_{X} F^{k0} \rangle \ud S_k \;.
\end{align}
This is the main result of this section. It shows that the Noether
charges associated with \textit{rigid} spatial symmetries can be
represented as flux integrals when the fields under consideration are
\textit{symmetric}.

It is instructive to see how the general Noether current
(\ref{Noether5}) assumes the special form (\ref{theta}) when the
symmetry conditions (\ref{symX}) are imposed. One has
\begin{align}
\label{T132}
\Theta^\mu=
X^{\alpha} T^{\mu}_{\alpha} & = - X^{\alpha} \langle F^{\mu\sigma}
F_{\alpha \sigma} \rangle + \frac{1} {2} X^{\alpha} \big(
(\mathcal{D}^{\mu} \Phi)^{\dagger} \mathcal{D}_{\alpha} \Phi +
(\mathcal{D}_{\alpha} \Phi)^{\dagger} \mathcal{D}^{\mu} \Phi \big) -
X^{\mu} \CL \;.
\end{align}
Using Eqs.(\ref{var}), (\ref{vvar}), and (\ref{WWW}), one obtains 
\begin{alignat}{2}
\label{ident}
F_{\sigma\mu} X^{\mu} & = D_{\sigma} \Psi_X \;, & \quad 
X^{\mu} \mathcal{D}_{\mu} \Phi & = \Psi_X \Phi \;.
\end{alignat}
As a result, the first term in
(\ref{T132}) can be transformed as
\begin{align}
- X^{\alpha} \langle F^{\mu \sigma} F_{\alpha \sigma} \rangle & =
- \langle F^{\sigma \mu} D_{\sigma} \Psi_X \rangle = - \langle D_{\sigma}
(F^{\sigma \mu} \Psi_X) \rangle + \langle \Psi_X D_{\sigma} F^{\sigma \mu}
\rangle \nonumber \\
& = - \partial_{\sigma} \langle \Psi_X F^{\sigma\mu} \rangle + \frac{1}
{2} \big(\Phi^{\dagger} \Psi_X \mathcal{D}^{\mu} \Phi - (\mathcal{D}^{\mu}
\Phi)^{\dagger} \Psi_X \Phi \big) \;,      
\end{align}
where the equations of motion have been used. Inserting this into
Eq.(\ref{T132}) and using Eq.(\ref{ident}) the terms containing the Higgs
field exactly cancel, giving
\begin{align}
\Theta^\mu=X^{\alpha} T^{\mu}_{\alpha} & = - \partial_{\sigma} 
\langle \Psi_X F^{\sigma\mu} \rangle - X^{\mu} \CL \;,   
\end{align}
which coincides with Eq.(\ref{theta}). 

\section{Calculation of the angular momentum}

Let us now choose $X = \partial_{\varphi}$ in Eqs.(\ref{M}),(\ref{Main}). 
This gives the conserved and gauge invariant
angular momentum
\begin{alignat}{2}
\label{JJJ}
J &=\int T^0_\varphi\, d^3x 
= - \oint \langle (A_{\varphi} - W_{\varphi}) F^{k0} \rangle \ud
S_{k} \;.
\end{alignat}
Here the second equality on the right applies 
for fields subject to the symmetry
conditions (\ref{symmetry}), $W_{\varphi}$ being the compensating
parameter in these conditions. In addition, one has to make sure that,
when transforming the volume integral into the surface integral, the
contribution of the inner boundary is zero.
This can be checked in the
gauge (\ref{s}), where $W_{\varphi} = 0$ while $A_{\varphi}$ given by
Eq.(\ref{axe}) is finite at the origin, so that the integral over a small
surface enclosing the origin would be nonzero only if the electric field
was $\sim 1/r^2$. This, however, would imply that the total energy is
infinite.
 
The surface integral structure of $J$ shows that only the asymptotic
long-range tails of the fields can contribute to the angular momentum.
In order to calculate this integral, it suffices therefore to analyze
the asymptotics of the fields near spatial infinity, where the problem
reduces to studying the linearized field equations. More precisely, let
$(A_{\mu}, \Phi)$ be a given static soliton solution with $J = 0$. We
consider all possible axial deformation of this solution with the only
condition that, asymptotically, the deformed configurations approach the
initial static solution, such that they will belong to the same
topological sector. Therefore, the deformed configurations can be
described by $(A_{\mu} + \psi_{\mu}, \Phi + \phi)$, where the
deformations $(\psi_{\mu}, \phi)$ can be arbitrary, with the only
condition that they vanish as $r \to \infty$. As a result, in the
asymptotic region the deformations satisfy the YMH equations linearized
around the $(A_{\mu},\Phi)$ background:
\begin{align}
\label{leq1}
D_{\sigma} D^{\sigma} \psi_{\mu} & - D_{\mu} D_{\sigma} \psi^{\sigma} +
2 [F_{\mu \sigma}, \psi^{\sigma}] - \mathcal{M}_{ab} \,\psi_{\mu}^{a}
\T_{b} \nonumber \\
& = \frac{1} {2} \big\{ \phi^{\dagger} \T_{a} \mathcal{D}_{\mu} \Phi -
(\mathcal{D}_{\mu} \Phi)^{\dagger} \T_{a} \phi + \Phi^{\dagger} \T_{a}
\mathcal{D}_{\mu} \phi - (\mathcal{D}_{\mu} \phi)^{\dagger} \T_{a} \Phi
\big\} \T_{a} \;, \\
\label{leq2}
\mathcal{D}_{\sigma} \mathcal{D}^{\sigma} \phi & + D_{\sigma} \psi^{\sigma}
\Phi + 2 \psi_{\sigma} \mathcal{D}^{\sigma} \Phi \nonumber \\
& = 
-\lambda \left\{(\Phi^{\dagger} \Phi - 1) \phi + (\Phi^{\dagger} \phi +
\phi^{\dagger} \Phi) \Phi \right\} \;,
\end{align}
where the mass matrix is
\begin{align}
\label{mass}
\mathcal{M}_{ab} & = \frac{1} {2} \Phi^{\dagger} (\T_{a} \T_{b} + \T_{b}
\T_{a}) \Phi \;.
\end{align}
Our strategy now is to solve these linearized equations in the
asymptotic region to see if there are modes giving a nonvanishing
contribution to the flux integral (\ref{JJJ}). We shall study axial
deformations of all known topological solutions for the gauge group
$\mathcal{G} = SU(2)$: 't~Hooft--Polyakov monopoles and Julia-Zee
dyons, sphalerons, and also vortices.

\subsection{'t~Hooft--Polyakov monopoles and Julia-Zee dyons}

These are spherically symmetric solutions of YMH theory with
$\mathcal{G} = SU(2)$ and the Higgs field in the real triplet
representation 
\cite{'tHooft:1974qc,Polyakov:1974ek,Julia:1975ff}. 
The gauge group generators
$\T_a$ are chosen according to Eq.(\ref{triplet}), $(\T_{a})_{ik} =
-\varepsilon_{aik}$. The mass matrix (\ref{mass}) has one zero
eigenvalue corresponding to a massless gauge boson. Hence, there are
long-range gauge field modes that may give a nonzero contribution to
Eq.(\ref{JJJ}).

Static, spherically symmetric YMH fields are characterized in this case
by the following gauge connection and the Higgs field (passing in the
gauge (\ref{s}) to spherical coordinates):
\begin{align}
\label{JZ}
A & = \Omega(r) \,\T_{3} \, \ud t + w(r)\left(-\T_{2} \, \ud \vartheta +
\T_{1} \sin \vartheta \, \ud \varphi \right) + \T_{3} \cos \vartheta \,
\ud \varphi \;,\\
\Phi^{k} & = \delta^{k}_3 \,\Phi(r) \;.
\end{align}
The field equations (\ref{eq1}) and (\ref{eq2}) reduce to 
\begin{align}
(r^2 \Omega')' & = 2 w^2 \Omega \;, \\
(r^2 \Phi')' & = 2 w^2 \Phi + \lambda r^2 (\Phi^2 - 1) \Phi \;, \\
r^2 w'' & = w (w^2 - 1) + r^2 (\Phi^2 - \Omega^2) w \;. 
\end{align}
The 't~Hooft--Polyakov monopoles ($\Omega = 0$) and Julia-Zee dyons
($\Omega \neq 0$) are solutions of this system that are regular at the
origin, corresponding to $\Omega(0) = \Phi(0) = 0$ and $w(0) = 1$, while
for large $r$  they approach exponentially fast
(for $\lambda \neq 0$) the asymptotic
values 
\begin{alignat}{3}
\label{asym}
\Omega & = \Sigma + \frac{Q} {r} \;, & \quad \Phi & = 1 \;, & \quad w &
= 0 \;,
\end{alignat} 
with constant $Q$, $\Sigma$. These solutions have
finite energy, electric charge $Q$, and unit magnetic charge. For
nonzero values of the self-coupling $\lambda$ these solutions can be
obtained numerically. For $\lambda = 0$ the Higgs field is massless and
has a long-range Coulomb tail: $\Phi = 1 + O(1/r)$ as $r \to \infty$. In
this case, the solution is known analytically
\cite{Bogomolny:1976de}:
\begin{alignat}{3}
\Omega & = \Sigma \Phi \;, & \quad 
\Phi & =\coth C r - \frac{1} {Cr} \;, & \quad
w & = \frac{Cr} {\sinh Cr},
\end{alignat}
with $C = \sqrt{1 - \Sigma^2}$. 

We would now like to study all possible axial deformations of these
solutions in the asymptotic region by solving the linearized equations
(\ref{leq1}) and (\ref{leq2}). The first step is to carry out a
multipole decomposition of perturbations to identify the most general
modes corresponding to axial deformations of the background solutions.
Since the backgrounds are spherically symmetric, the angular quantum
number $j$ is conserved and perturbations for different values of $j$
decouple from each other.  It is convenient to introduce the basis of
complex one-forms
\begin{align}
\theta^0 & = \ud t \;, & \theta^1 & = \ud r \;, & \theta^2 & = \frac{r}
{\sqrt{2}} (\ud \vartheta - \im \sin \vartheta \, \ud \varphi) \;, &
\theta^3 & = (\theta^2)^\ast \;,
\end{align}
whose nonzero scalar products $\theta^{\alpha\beta} \equiv
(\theta^{\alpha},\theta^{\beta})$ are $\theta^{00} = -\theta^{11} =
-\theta^{23} = 1$. In addition, one introduces the new Lie algebra
basis $\textrm{L}_{1} = \T_{1} + \im \T_{2}$, $\textrm{L}_{2} = \T_{1}
- \im \T_{2}$, $\textrm{L}_{3} = \T_{3}$.  The perturbations are then
expanded as
\begin{align}
\psi_{\mu} \ud x^{\mu} & = \textrm{L}_{a} \psi_{\alpha}^{a}
\theta^{\alpha} \;, & 
\phi^{a} & = \langle \textrm{L}_{a} \T_{b} \rangle f^{b} \;.
\end{align}
A complete separation of the angular variables in the perturbation
equations (\ref{leq1}) and (\ref{leq2}) is achieved by making the
following ansatz:
\begin{align}
\psi^{a}_{\alpha} & = Z^{a}_{\alpha}(r) \, _{s}Y_{jm}(\vartheta,\varphi)
\;, & 
f^{a} & = U^{a} (r) \, _{\sigma}Y_{jm}(\vartheta,\varphi) \;.
\end{align}
Here, $_{s}Y_{jm} (\vartheta,\varphi)$ are the spin-weighted spherical
harmonics \cite{Goldberg:1967uu}. 
The quantum numbers $j$, $m$ are the same for
all values of the indices $a$,$\alpha$, while the values of the spin
weights $s = s(a,\alpha)$ and $\sigma = \sigma(a)$ are determined by
direct inspection of Eqs.(\ref{leq1}) and (\ref{leq2}) using
the properties of the  spin-weighted harmonics \cite{Goldberg:1967uu}.

Within the multipole decomposition obtained, we specialize to the dipole
($j = 1$) and axially symmetric ($m = 0$) sector.  The most general
perturbations in this case are described by (passing back to the
standard basis)
\begin{align}
\label{j=1}
\psi & = \left(\T_{1} \frac{Z_{1} (r)} {r} \sin \vartheta + \T_{3}
\frac{Z_{2} (r)} {r} \cos \vartheta \right) \ud t  + \T_{2} \, Z_{3} (r)
\sin \vartheta \, \ud r + \T_{2} \, Z_{5} (r) \cos \vartheta \, \ud
\vartheta \nonumber \\
& \hphantom{ = } + \left(-\T_{1} \, Z_{5} (r) \cos \vartheta + \T_{3} \,
Z_{4} (r) \sin \vartheta \right) \, \sin \vartheta \, \ud \varphi \;,
\nonumber \\
\phi^{k} & = \delta^{k}_{1} \frac{U_{1} (r)} {r} \sin \vartheta +
\delta^{k}_{3} \frac{U_{2} (r)} {r} \cos\vartheta \;.
\end{align}
This ansatz has a residual U(1) gauge symmetry generated by the 
infinitesimal gauge transformations (\ref{3}) with $U = \exp(-L)$,
\begin{alignat}{2}
\psi & \to \psi + \ud L + [A,L] \;, & \quad \phi & \to \phi - L\Phi \;,
\end{alignat}
where $L = \alpha(r) \T_{2} \sin \vartheta$. This symmetry does not
change the values of $Z_{2}$ and $U_{2}$, while
\begin{align}
Z_{1} & \to Z_{1} - r \Omega \alpha \;, &
Z_{3} & \to Z_{3} + \alpha' \;, &
Z_{4} & \to Z_{4} + w \alpha \;, &
Z_{5} & \to Z_{5} + \alpha \;, &
U_{1} & \to U_{1} - r \alpha \Phi \;,
\end{align}
which can be used to impose the gauge condition\footnote{There remains
one pure gauge mode generated by constant $\alpha$.} $Z_3 = 0$.
Inserting now the ansatz (\ref{j=1}) into the perturbation equations
(\ref{leq1}) and (\ref{leq2}), the angular dependence decouples and we
obtain a system of radial equations for the amplitudes $Z_1$, $Z_2$,
$Z_4$, $Z_5$, $U_1$, $U_2$ which is listed in the Appendix. Inserting
the ansatz into the angular momentum integral (\ref{JJJ}) gives (we are
working in the gauge (\ref{s}) where $W_{\varphi} = 0$)
\begin{align}
J & = \lim_{r \to \infty} r^2 \oint \langle (A_{\varphi} +
\psi_{\varphi}) (A_{0} + \psi_{0})^{\prime} \rangle \sin \vartheta \,
\ud \vartheta \, \ud \varphi \nonumber \\
& = \frac{4 \pi} {3} \lim_{r \to \infty} r^2 \left(2 w \left(\frac{Z_1}
{r} \right)^\prime + \left(\frac{Z_2} {r}\right)^\prime + 2
\Omega^\prime Z_4 \right) \;.
\end{align}
Since the background amplitudes approach their asymptotic values 
(for large $r$) exponentially fast, we can replace $\Omega$, $\Phi$, $w$ by
their asymptotics  (\ref{asym}). This gives
\begin{align}
\label{fin}
J & = \frac{4\pi} {3} \lim_{r \to \infty} r^2 \left(
\left(\frac{Z_2} {r} \right)^\prime - \frac{2 Q} {r^2} Z_4 \right) \;.
\end{align}
The asymptotic behavior of the amplitudes $Z_2$ and $Z_4$ is determined
from the radial equations (\ref{A2}) and (\ref{A3}), which in the
asymptotic region reduce to
\begin{alignat}{2}
\left(-\frac{\ud^2} {\ud r^2} + \frac{2} {r^2} \right) Z_2 & = 0 \;, &
\quad
\left(-\frac{\ud^2} {\ud r^2} + \frac{2} {r^2} \right) Z_4 & = 0 \;. 
\end{alignat}
Solutions that are regular at infinity are
\begin{align}
Z_2 & \sim \frac{1} {r} \;, & Z_4 & \sim \frac{1} {r} \;. 
\end{align}
Inserting these into Eq.(\ref{fin}) finally gives\footnote{The same result
is obtained for $\lambda = 0$, in which case  all
perturbation equations can be solved exactly \cite{Heusler:1998ec}.}
\begin{align}
J & = 0 \;.
\end{align} 
In fact, in order to ensure a nonzero value of $J$, the amplitudes
$Z_2$, $Z_4$ should approach nonzero constant values at infinity, which
is not the case. The conclusion is that there are no stationary, axial
deformations of the 't~Hooft--Polyakov monopoles and Julia-Zee dyons that
would support a nonzero angular momentum. The same is true for higher
(quadrupole, etc.)  multipole deformations, since all of them decay at
infinity even faster than the dipole ones. This conclusion did not
require smallness of deformations for all $r$, the only requirement
having been that deformed configurations must approach the spherically
symmetric solutions for $r \to \infty$%
\footnote{
The rotational excitations of monopoles were also studied  
in Ref.\cite{Heusler:1998ec}; this work, however, used 
the {\it volume} integral 
representation of the angular momentum. In view of this,  it was necessary 
to assume the perturbative regime of rotational deformations {\it everywhere},
thus restricting consideration
to the case of {\it slow} rotation. In our analysis,
on the other hand, the rotation is not assumed to be slow.}.

\subsection{Sphalerons and vortices}

Sphalerons are spherically symmetric solutions of a YMH theory with
$\mathcal{G} = SU(2)$ and the Higgs field in the complex doublet
representation \cite{Klinkhamer:1984di,Kunz:1989sx}.  
The gauge group generators
$\T_{a}$ are thus chosen according to Eq.(\ref{dublet}), $(\T_{a}) =
\frac{1} {2 i} \tau^{a}$. In the simplest case
\cite{Klinkhamer:1984di}, static and spherically symmetric YMH fields are
characterized by the following purely magnetic gauge connection and
Higgs field:
\begin{align}
\label{JZa}
A & = w(r) \left(-\T_{2} \, \ud \vartheta + \T_{1} \sin \vartheta \, \ud
\varphi \right) +\T_{3} \cos\vartheta \, \ud \varphi \;,\\
\Phi^{k} & = \delta^{k}_{1} \, \Phi(r) \;.
\end{align}
The field equations (\ref{eq1}) and (\ref{eq2}) reduce to 
\begin{align}
(r^2 \Phi')' & = \frac{1} {2} (w + 1)^2 \Phi + \lambda r^2 (\Phi^2 - 1)
\Phi \;, \\
r^2 w'' & = w(w^2 - 1) + \frac{r^2} {2} \Phi^2 (w + 1) \;. 
\end{align}
Sphalerons are solutions of this system which are regular at the
origin ($\Phi(0) = 0$, $w(0) = 1$) and approach the asymptotic values
\begin{alignat}{2}
\label{asym1}
w & = -1 \;, & \quad \Phi & = 1 \;
\end{alignat}
for large $r$ exponentially fast. The crucial point now is that all
deformations of these background solutions also approach zero exponentially
fast. This is a manifestation of the fact that the gauge symmetry of
the vacuum (\ref{asym1}) is broken completely, since all eigenvalues of
the mass matrix (\ref{mass}) are nonzero. As a result, there are no
long-range solutions of the linearized field equations, and the angular
momentum integral is zero. The only subtlety is the limit $\lambda \to
0$, since then the Higgs field becomes long range. However, as the
background fields are purely magnetic, the equations for the most
general dipole, axially symmetric gauge field perturbations do not
contain any Higgs field perturbations\footnote{The same thing happens
for the dyons, since Eqs. (\ref{A1}) and (\ref{A2}) decouple from the
rest in the purely magnetic limit $\Omega\to 0$.}.
The relevant perturbation equations, therefore, contain only massive
amplitudes. Thus, their solutions approach zero exponentially fast. The
conclusion\footnote{This conclusion also applies to the deformed
sphalerons of \cite{Kunz:1989sx}.} is that there are no stationary and
axially symmetric  spinning excitations of sphalerons.

To complete our considerations, we also want to consider the YMH vortices.
It is known that the Abelian Nielsen-Olesen vortex \cite{Nielsen:1973cs} 
does
not admit spinning generalizations within the original YMH theory with
$\mathcal{G} = U(1)$ \cite{Julia:1975ff}. 
However, it is not excluded that such
generalizations may exist within a YMH theory with a larger gauge group
$\mathcal{G}$. Let us restrict consideration to cylindrically symmetric, i.e.,
$z$-independent, YMH fields. Then one can straightforwardly obtain from
Eq.(\ref{JJJ}) the angular momentum per unit length $z$,
\begin{align}
\label{JJJJ}
J & = -\oint \langle (A_{\varphi} - W_{\varphi}) F_{0 \rho} \rangle \,
\ud l \;,
\end{align}
where the integration is over a circle of radius $\rho \to\infty$ in a
plane of constant $z$. For spinning excitations that asymptotically
approach the Nielsen-Olesen vortex, both $A_{\varphi}$ and $W_{\varphi}$
stay finite as $\rho \to \infty$, and so the integral will be nonzero
if only $F_{0\rho} \sim 1/\rho$. However, this would imply that the
energy is divergent. The conclusion is that there are no
axially symmetric, spinning excitations of the Nielsen-Olesen vortex
within YMH theory\footnote{Spinning vortices can exist in generalized YMH
theories including the Chern-Simons term 
\cite{deVega:1986eu,Jackiw:1990aw}.}  for a compact gauge group 
$\mathcal{G}$. 

\section{Concluding remarks}

Summarizing our results, we have shown that none of the ``canonical'' 
topological solitons of the $\mathcal{G}=SU(2)$ YMH theory admit 
spinning excitations in the stationary and axisymmetric one-soliton sector. 
Although not completely eliminating all spinning
solitons in gauge field theory, this conclusion renders their existence  
somewhat less probable. 
Therefore, we would like to list
the remaining possibilities for constructing spinning solutions
(if they exist at all). First, one can try to consider YMH theories with
$\mathcal{G}> SU(2)$, which might work in the case of monopoles or dyons.
The pattern of symmetry breaking can be quite different for higher gauge
groups and for different representations of the Higgs field. If there
remain several massless gauge group generators after symmetry breaking,
then there is a better chance to have long-range modes giving
a contribution to the angular momentum surface integral%
\footnote{In the Einstein-Yang-Mills theory, for example,
where the symmetry is not broken at all, there exist static solitons whose
linear axial deformations do support a nonzero angular momentum 
\cite{Brodbeck:1997ek}.
It is, however, unclear at present whether these linear rotational 
modes can be promoted to spinning solutions also at the nonlinear 
level \cite{VanderBij:2001nm}. 
}.

The other possibility is to consider YMH systems that are \textit{not}
symmetric under the combined action of axial rotations and gauge
transformations, while their action \textit{is} symmetric. The angular
momentum then will still be conserved, but it will be given by a volume
integral. Thus, it may receive contributions also from short-range
field modes.

Finally, we would like to make some remarks on the nonexistence of
rotating monopoles. First, it should be emphasized 
that monopoles do not rotate only within classical theory. Quantum
monopoles, on the other hand, \textit{do} have angular momentum
associated with the fermionic zero modes \cite{Jackiw:1976fn}; this
effect, however, disappears in the classical limit. 
For example, supersymmetric monopoles are conjectured to be
dual to the elementary particles with spin (Monteon-Olive duality), 
thus implying that monopoles themselves have a spin. 
However, this spin is carried by the fermionic superpartners
of monopoles and not by the bosonic monopole configurations. 

Second, it is
well known that the angular momentum of an electric charge
moving around a magnetic monopole contains an extra term that can be
interpreted as the angular momentum of the field \cite{Jackiw:1980ub}.
At first glance, this disagrees with our conclusion that the angular
momentum of the monopole field is zero. However, this extra term does
not in fact relate to the monopole alone, but to the system of both
charges, one of which is electric and the other magnetic. Even when
these charges are at rest, the angular momentum of the total field $\int
\vec{r}\times(\vec{E}\times\vec{B}) \, \ud^3 x$ does not vanish.
However, if the electric field $\vec{E}$ of the electric charge is zero
(no charge), the contribution of the magnetic charge alone will be zero.

We would also like to emphasize once again that our results apply
only within the {\it one-soliton} sector, thus showing the absence of 
{\it spinning} excitations of isolated solitons. Outside this sector
one can have solutions with $J\neq 0$ describing {\it orbital} motions 
of solitons. Such solutions are explicitly known in the case of rotating 
monopole-antimonopoles pairs  
\cite{Heusler:1998ec,VanderBij:2001nm,Kleihaus:1999sx}%
\footnote{However, the axially symmetric dyons with higher values 
of topological charge
\cite{Hartmann:ja} do not rotate 
\cite{VanderBij:2001nm}.}. It is also not excluded that in many-soliton
systems, as for example in soliton scatterings,  solitons might develop
some kind of  spinlike deformation due to their mutual polarization. 
However, such deformations will tend to zero in the 
limit of infinite separation of solitons.

\section{Acknowledgments}

M.S.V. thanks Peter Forgacs for numerous discussions
and acknowledges conversations with Bernard Julia,  
Jochum van der Bij, and Eugen Radu. 
E.W. thanks Tom Heinzl for many discussions and a
careful reading of the manuscript. The work of E.W. was supported by the
DFG. We would also like to thank Andreas Wipf for
his support and comments.  

\section*{Appendix}

\renewcommand{\theequation}{A.\arabic{equation}}

\setcounter{equation}{0}

In this appendix we list the full system of radial equations describing
the most general stationary, axially symmetric excitations of the
Julia-Zee dyons. These equations are obtained by putting Eq.(\ref{j=1})
(with $Z_3 = 0$) into the field equations (\ref{leq1}) and (\ref{leq2}),
\begin{align}
\label{A1}
0 & = \left(-\frac{\ud^2} {\ud r^2} + \frac{w^2 + 1} {r^2} + \Phi^2
\right) Z_1 - \frac{2w} {r^2} Z_2 + \frac{\Omega} {r} (Z_5 - w Z_4) -
\Omega \Phi \, U_1 \;, \\
\label{A2}
0 & = \left(-\frac{\ud^2} {\ud r^2} + 2 \frac{w^2 + 1} {r^2} \right) Z_2
- \frac{4w} {r^2} Z_1 - \frac{4 w \Omega} {r} Z_5 \;, \\
\label{A3}
0 & = \left(-\frac{\ud^2} {\ud r^2} + \frac{w^2+2} {r^2} \right) Z_4 -
\frac{3w} {r^2} Z_5 + \frac{w} {r} (\Omega Z_1 - \Phi U_1) \;, \\
\label{A4}
0 & = \left(-\frac{\ud^2} {\ud r^2} + \frac{3 w^2} {r^2} + \Phi^2 -
\Omega^2 \right) Z_5 - \frac{3w} {r^2} Z_4 + \frac{\Omega} {r} (2 w Z_2
- Z_1) + \frac{\Phi} {r} (U_1 - 2 w U_2) \;, \\
\label{A5}
0 & = \left(-r \Omega \frac{\ud}{\ud r} + (r \Omega)' \right) Z_1 -
\frac{\ud Z_5} {\ud r} + \left(-w \frac{\ud} {\ud r} + w' \right) Z_4 +
\left(r \Phi \frac{\ud} {\ud r} - (r \Phi)' \right) U_1 \;, \\
\label{A6}
0 & = \left(-\frac{\ud^2} {\ud r^2} + \frac{w^2 + 1} {r^2} - \Omega^2 +
\lambda(\Phi^2 - 1) \right) U_1 - \frac{2w} {r^2} U_2 + \frac{\Phi} {r}
(Z_5 - w Z_4) + \Omega \Phi Z_1 \;, \\
\label{A7}
0 & = \left(-\frac{\ud^2} {\ud r^2} + 2 \frac{w^2 + 1} {r^2} + \lambda(3
\Phi^2 - 1) \right) U_2 - \frac{4w} {r^2} U_1 - \frac{4 w \Phi} {r} Z_5
\;.
\end{align}
It is instructive to verify that for $\lambda = 0$ these equations admit
a global symmetry: if $\{Z_1(r), Z_2(r), Z_4(r), Z_5(r), U_1(r),
U_2(r)\}$ is a solution for the purely magnetic background
$\{\Omega(r)=0, \Phi(r), w(r)\}$, then
\begin{align*}
{Z}^\gamma_1(r) & = Z_1(\gamma r) + \sqrt{1 - \gamma^2} U_1(\gamma r)
\;, & 
{Z}^\gamma_2(r) & = Z_2(\gamma r) + \sqrt{1 - \gamma^2} U_2(\gamma r)
\;, \\
{U}^\gamma_1(r) & = U_1(\gamma r) + \sqrt{1 - \gamma^2} Z_1(\gamma r)
\;, & 
{U}^\gamma_2(r) & = U_2(\gamma r) + \sqrt{1 - \gamma^2} Z_2(\gamma r)
\;, \\
{Z}^\gamma_4(r) & = \gamma Z_4(\gamma r) \;, &
{Z}^\gamma_5(r) & = \gamma Z_5(\gamma r)
\end{align*}
is a solution corresponding to a ``rotated'' background characterized by  
\begin{alignat*}{3}
{\Omega}^\gamma(r) & = \sqrt{1 - \gamma^2} \Phi(\gamma r) \;, & \quad
{\Phi}^\gamma(r) & = \Phi(\gamma r) \;, & \quad
w^\gamma(r) & = w(\gamma r) \;.
\end{alignat*}

\end{document}